\documentclass[prl,aps,twocolumn,showpacs]{revtex4}
\usepackage{graphicx}
\usepackage{dcolumn}
\usepackage{amssymb}
\usepackage{amsmath}

\begin{document}

\title{The onset of void coalescence during dynamic fracture
of ductile metals}

\author{E.\ T.\ Sepp\"al\"a}
\author{J.\ Belak}
\author{R.\ E.\ Rudd}

\affiliation{Lawrence Livermore National Laboratory, Condensed Matter Physics 
Division, L-415, Livermore, CA 94551, USA}
\date{\today}

\begin{abstract}
Molecular dynamics simulations in three-dimensional copper have been
performed to quantify the void coalescence process leading to
fracture. The correlated growth of the voids during their linking is
investigated both in terms of the onset of coalescence and the ensuing
dynamical interactions through the rate of reduction of the distance
between the voids and the directional growth of the voids.  The
critical inter-void ligament distance marking the onset of coalescence
is shown to be approximately one void radius in both measures.
\end{abstract}


\maketitle

The point at which voids begin to coalesce during dynamic fracture is
of considerable interest because complete fracture of the material
typically ensues rapidly thereafter.  A robust model of the onset of
coalescence is an important ingredient of a predictive model of
dynamic fracture.  The conventional picture of how ductile metals
break under the rapid application of stress consists of three stages:
void nucleation, growth and coalescence.  Initially voids nucleate
from the weak points in the material such as inclusions and/or grain
boundary junctions.  Once nucleated, the voids grow under the tensile
stress, driven by the reduction in elastic energy.  Eventually, the
voids grow sufficiently large that they interact with each other,
coalesce into larger voids, and finally form the fracture
surface~\cite{Curran,McClintock}.  There are many interesting twists
and subtleties, such as the interplay between shear localization and
void growth, but the basic picture applies to the fracture of a broad
class of ductile metals.  Naturally considerable effort has gone into
the study of this fracture process, both in
modeling and theory and in experiment, 
including a new generation of 3D, 
non-destructive
fracture characterization techniques such as x-ray tomography.
Nevertheless, a robust, mechanistic
understanding of coalescence has yet to emerge.

Computationally void growth has been studied extensively at the
continuum level,~\cite{RiceTracey,McClintock2,Gurson,tvergaard} and
more recently at the atomistic level~\cite{belak,rudd,prb}.  The
atomistic studies demonstrate that voids grow by emitting
dislocations, which carry away the material, platelets of atoms, from
the void and are responsible for the plastic deformations needed to
accommodate significant void growth.  There are also many recent
studies of fracture in ductile metals with several holes or
voids~\cite{koss}. 
While these studies model the void growth explicitly, often with fairly
sophisticated models of plasticity, they typically
simplify the coalescence process to instantaneous unification of
the voids based on a relatively simple criterion such as growth
of the voids to within one void diameter of each other or a
plastic strain threshold.
The earlier continuum studies (cf.\ Ref.~\cite{tvergaard}) 
and the one atomistic study
known to us~\cite{somerday} of the coalescence process have been
conducted in effectively two dimensional and highly symmetric systems.

In this Letter we analyze the details of the onset of void
coalescence.  In particular we quantify the point at which coalescence
begins, as measured by a critical {\it inter-void ligament distance}
(ILD), and examine the mechanisms involved in the transition from
independent void growth to coalescence.  There are several ways in
which two voids can interact.  In the case of pure impingement, the
voids only interact when they grow to the point that they intersect
and join into a single void.  In reality, the voids interact before
they intersect.  Their range of interaction is extended due to their
elastic and plastic fields.  Each void generates an elastic strain
field of the form generally associated with centers of dilatation
\cite{Eshelby}. The shear stress decreases with the distance from the
void like $r^{-3}$.  For voids sufficiently close each void's growth
rate is altered by the stress field of the proximal void.  The
modification of the elastic field can affect the initiation of
plasticity, as well as the subsequent development of the plastic zone
around the voids. The voids may interact through their plastic fields,
too, in which case the fields may give rise to an increased hardening
rate in a localized region or to thermal softening and shear
localization.  An argument due to Brown and Embury for a transition to
shear deformation based on simple geometrical considerations suggests
that the critical inter-void ligament distance, ILD$_c$, should be
equal to one diameter of a void~\cite{Brown}; that is, when the
surfaces of a pair of voids are separated by one void diameter, they
transition from independent void growth to coalescence.  It is at this
point, they argue, that the dominant void process switches from the
radial plastic flow around isolated growing voids to a shear
deformation allowing the rapid coalescence of the pair of voids.
However, more recent two-dimensional studies suggest that for
distances between voids as large as six diameters the void growth rate
is enhanced~\cite{horstemeyer}.

The use of atomistic techniques permits the analysis of the
contributions of these competing mechanisms to the onset of void
coalescence, as we describe in this Letter. We demonstrate the
existence of, and compute, the critical inter-void ligament distance
ILD$_c$ by starting with two voids well separated from each other and
detecting the point at which correlated growth begins, marked both by
the accelerated rate at which the two void surfaces approach each
other and by biased growth causing the voids to start to extend toward
each other. This gives an indication of the onset of the coalescence
process, and it tests the argument by Brown and Embury~\cite{Brown}.
We also test the setup by Horstemeyer {\it et al}~\cite{horstemeyer}
by varying the initial distances between the voids and measuring the
asymptotic growth rate of the voids.  The initial void-to-void
distance below which the growth-rate is enhanced should give another
candidate for the critical distance and measure it in a volumetric
sense.

\begin{figure}
\includegraphics[width=75mm]{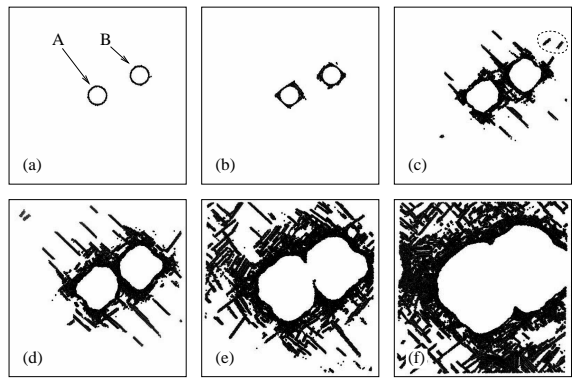}
\caption{Snapshots of the slices of the two-void system at
$\dot{\varepsilon} = 10^9$/sec with only those atoms shown that are in
dislocation cores, stacking faults, void surfaces or other defects
(see text). The dashed loop in panel (c) is drawn around a slice of a
prismatic dislocation loop.  The plane shown passes through the
centers of both voids.  The snapshots show the initial plasticity
(a,b), interacting plastic zones (c,d) and the final coalescence
(e,f).  The frames correspond to strains of
$\varepsilon = 1.72\%$, 2.42\%, 3.47\%, 3.89\%, 4.52\%, and 5.21\%,
respectively.
}
\label{fig1}
\end{figure}

We have performed large-scale (parallel) classical molecular dynamics
(MD) simulations~\cite{allen} in single crystal face-centered cubic
(FCC) systems using an empirical embedded-atom model (EAM) potential
for copper~\cite{oh1}.  The three dimensional (3D) simulation box
consists of $120\times120\times120$ 4-atom FCC unit cells with 
periodic boundary
conditions for a total of 6~912~000 atoms.  The system is initially
equilibrated using a thermostat~\cite{hoover} at $T=300$~K and a
constant volume $L^3$ (with $L=43.4~\mbox{nm}$) chosen to give ambient
pressure, $P \simeq 0$~MPa.  Once the system has reached equilibrium,
two spherical voids are cut in the system with radius $r_0 = 0.05 L =
2.2~\mbox{nm}$: one in the middle of the box and the other 12.2~nm
away, in a relative position of $[0.25, 0.1166, 0.0544]L$.  We refer
to these as void A and void B, respectively, see Fig.~\ref{fig1}. When
the initial distance between the voids is varied, the location of the
void B is changed, but the relative orientation of the voids is kept
fixed.  Initially, the voids are equal in size, with approximately
3620 atoms removed for each.  Once the voids are formed, the
thermostat is turned off, and dilatational strain is applied uniformly
at a constant strain-rate $\dot{\varepsilon}$.  Applied strain-rates
of $\dot{\varepsilon}=10^8$/sec and $10^9$/sec have been used with
perfectly triaxial, or hydrostatic, expansion.  The dilatation is
simulated by expanding the simulation box at each time
step~\cite{belak}, implemented in the fashion of the Parrinello and
Rahman technique~\cite{parrinello} in order to prevent the spurious
generation of elastic waves at the box boundaries.  More details of
the simulation method can be found from Ref.~\cite{prb}.

While some void growth takes place through elastic stretching in the
initial phases of the box expansion, significant void growth and
void-void interaction take place only once plastic deformation has
begun.  The important role of plasticity forces us to consider in some
detail how dislocations are generated and the effect the dislocation
dynamics have on void coalescence.  Figure~\ref{fig1} shows a
visualization within a slice of width 4.5~{\AA} of a plane including
centers of both voids at six different instants during coalescence.
The atoms shown are either on the surface of the voids or belong to
dislocations. The decision of which atoms to plot is based on a
geometrical criterion, a finite-temperature generalization 
of the centrosymmetry
deviation~\cite{rudd,hamilton}.  From the snapshots one sees that the
deformation mechanism involves the nucleation and propagation of
dislocations, accommodating the void growth, and the interaction of
the dislocations.  For
example, the prismatic dislocation loops punched out by the voids
appear as roughly parallel line traces (due to the stacking fault
ribbons) in the slice Fig.~\ref{fig1}(c), as verified in the full 3D
configuration.  Initially the dislocation activity around each void is
essentially symmetric [Fig.~\ref{fig1}(a) and (b)], as expected for
independent void growth, but as the plastic fields evolve the
void-void interaction is clearly evident both through interactions
between the two plastic zones and bias due to the elastic fields
[Fig.~\ref{fig1}(c)].  Once the dislocation density grows sufficiently
high in the ligament region between the voids [Figs.~\ref{fig1}(d)],
void B begins to grow in the direction away from void A.  Eventually
the voids coalesce [Fig.~\ref{fig1}(e)], and continue to grow as one
until ultimately coalescence with the void's periodic images takes
place [subsequent to Fig.~\ref{fig1}(f)] so that the cavity percolates
through the periodic system.

\begin{figure}
\includegraphics[height=8cm,angle=-90]{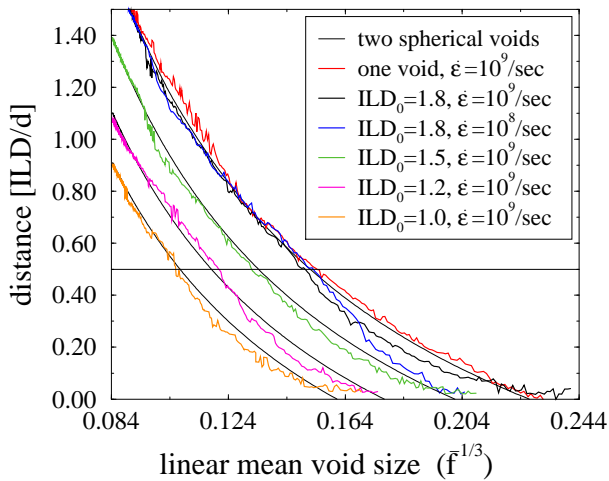}
\caption{ (color) The dynamical ILD, the distance between the surfaces
of the voids along the line connecting the original center positions,
plotted versus the mean void size $\bar{f}^{1/3}$ for various initial
ILD$_0$'s.  The ILD is plotted in units of the average void diameter
$d$ calculated at each instant using the formula $d=2 [3/(4 \pi)
\bar{f} V]^{1/3}$ assuming roughly spherical voids.  For reference,
thin lines are plotted to show the relationship for spherical voids
impinging freely on each other (see text).  The
red line shows the hypothetical ILD computed by duplicating the single
void at fixed centers with ILD$_0=1.8$.  
The horizontal line is at ILD=0.5$d$.}
\label{fig2}
\end{figure}

Figure~\ref{fig1} offers several visual indications of the interaction
between voids.  Clearly, the separation between the void surfaces (the
ILD) serves as something akin to a reaction coordinate for the
coalescence: the voids coalesce when it goes to zero.  Other
indications include the displacement of the center of a void as it
grows preferentially toward the neighboring void and the change in the
void growth rate as the voids interact.  We now quantify these effects
in order to analyze the coalescence.

In Fig.~\ref{fig2} the dynamic evolution of the ILD has been plotted
for strain-rates $\dot{\varepsilon}=10^8$/sec and $10^9$/sec and for
various {\it initial} closest surface-to-surface distances between the
voids ILD$_0$.  In Fig.~\ref{fig1} the case ILD$_0 \simeq$ 1.8 was shown.  
While the strain is the quantity controlled in these simulations, we
have found that the data collapse well and the coalescence is
indicated much more clearly if the ILD data are plotted as a function
of the linear mean void size, $\bar{f}^{1/3}$, where $\bar{f}$ is the
average of the two void fractions $f=V_{void}/V$ and $V$ is the
volume of the box. The technique for calculating the void volume is
described in Ref.~\cite{prb}. The reason is that for voids growing
independently, the ILD is closely related to the size of the void
through geometry.  In particular, the ILD for two spherical voids of
the same diameter $d$ growing independently is given by
ILD~=~(ILD$_0+d_0)(1+\varepsilon)-d$, where $d_0$ is the initial
diameter.  This formula is used to generate the thin curves in
Fig.~\ref{fig2}. Deviations from these free impingement curves in the
MD simulations indicate void shape changes, most importantly anisotropic
growth due to coalescence.  The same data plotted versus
strain exhibit a strong strain-rate dependence inherited from the void
growth law $V_{void}(\varepsilon)$ (cf. Ref.~\cite{prb}).
For reference, the snapshots in Fig.~\ref{fig1}(a)-(d) correspond to
mean linear void size $\bar{f}^{1/3} = 0.089$, 0.094, 0.150, and 
0.195, respectively.  After coalescence $\bar{f}$ is not calculated.

Initially the separation distance decreases essentially
smoothly until the plasticity begins, eventually reaching zero. A
transition occurs when the ILD starts to decrease much faster than the free
impingement line, which takes place when the ILD reaches
approximately one half: ILD$_c =0.5 \pm 0.1$ diameter or one radius,
independently of ILD$_0$ or the strain-rate.  Note that the unit of
ILD is the current diameter of a void, $d$, not the initial
value. A curve derived from a single void growth is
provided to estimate the contribution of uncorrelated faceting effects
(the ``one void'' curve at ILD$_0$=1.8), and these effects are seen to
be relatively small.  The critical ILD of one radius is much lower
than the Brown-Embury estimate, and it corresponds to a strain of
3.47\% ($\bar{f}^{1/3} \simeq 0.15$) for ILD$_0$=1.8 at
$\dot{\varepsilon}=10^9$/sec, corresponding to frame (c) of
Fig.~\ref{fig1}.  In the very final stages the ligament is drawn under
biaxial stress, and the flow switches from radial material transport
to tangential transport as the mechanism switches from loop punching
to drawing.  At this point, the material is highly defective but it
remains ductile.  There is no abrupt fracture, as might be expected at
larger length scales.  Coalescence results from extended
drawing and thinning of the ligament until rupture.

\begin{figure}
\includegraphics[height=8cm,angle=-90]{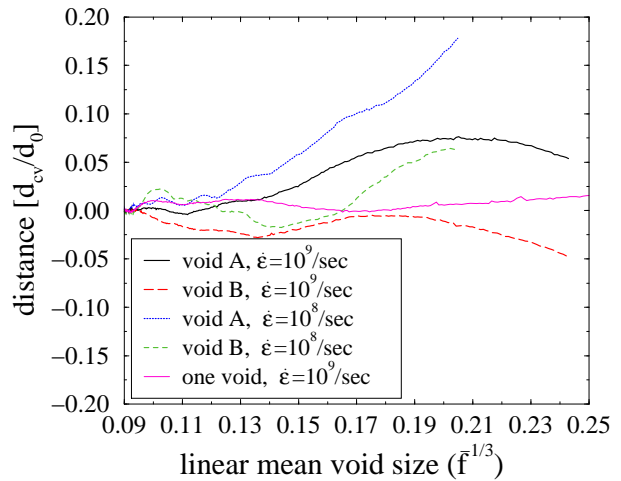}
\caption{Distance $d_{cv}$ from the original center of void to the
instantaneous void center, projected onto the line connecting the
original void centers, plotted versus the average void size to the
point of coalescence (ILD $\simeq 0$).  The sign of the distance
$d_{cv}$ is positive for motion toward the other void.  ILD$_0$=1.8.  
The thin solid line is
for a single void in the same size of the box and with the same radius
and $\dot{\varepsilon}=10^9$/sec projected to the same line. Here the
distance $d_{cv}$ is given in the units of the original void diameter
$d_0$.}
\label{fig3}
\end{figure}

Another measure of void interactions is whether the voids grow
preferentially toward their neighbor.  This effect is quantified in
Fig.~\ref{fig3}, which shows the motion of the center of mass of the
void surface (the void center) for the voids shown in Fig.~\ref{fig1}.
Here ILD$_0$=1.8 and $\dot{\varepsilon}=10^9$/sec.  After the void
growth starts, the center of void A initially moves only slightly, but
at about $\bar{f}^{1/3} = 0.15$ (ILD $=0.5$ in Fig.~\ref{fig2}), it
starts to move in the direction of the other void as the void growth
becomes biased toward its neighbor.  Just before coalescence the
center of void A begins to move away from void B, as the growth is
biased in the opposite direction. During this sequence, void B
initially grows away from void A, then roughly in unison with void A
($\bar{f}^{1/3} = 0.15$) it begins to grow toward its neighbor, and
before coalescence it too switches to growth away from the proximal
void. This retrograde growth happens at the same point (after
$\bar{f}^{1/3}=0.19$) as the decrease of ILD begins to slow down in
Fig.~\ref{fig2} [see also the snapshot in Fig.~\ref{fig1}(d)].  The
same phenomenon--first slow movement or repulsion from the void; then
growth toward the nearby void after $\bar{f}^{1/3} = 0.15$ (ILD
$=0.5$); and finally retrograde growth--holds in the
$\dot{\varepsilon}=10^8$/sec case, too, Fig.~\ref{fig3}. As a
reference the movement of the center of a single void in a box with
$\dot{\varepsilon}=10^9$/sec projected to the same line is plotted.
Comparing the single void case with the interacting voids at same
strain-rates, one sees that the maximum distance the centers of the
interacting voids have moved is three to five times larger than the
nanoscale random
walk of the single void center, so the movement of the void center is
not just due to statistical fluctuations at the void surface.

\begin{figure}
\includegraphics[height=8cm,angle=-90]{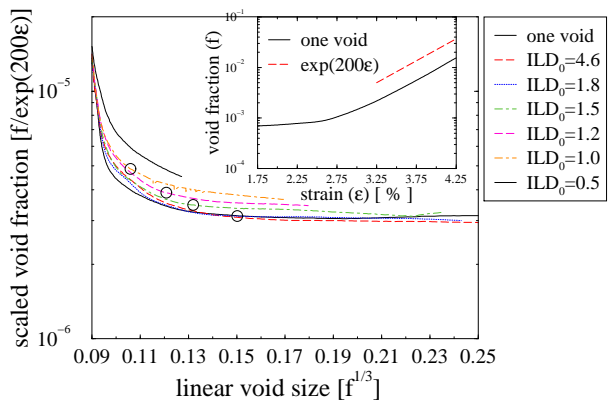}
\caption{Growth of the voids until coalescence presented by void
fraction for ILD$_0$=0.5, 1.0, 1.2, 1.5, 1.8, and 4.6 diameters as well
as in a single void case (in same box size) at $\dot{\varepsilon}
=10^9$/sec.  The asymptotic behavior (before finite size effects) is
exponential growth with $\exp(200\epsilon)$ as seen in the inset for
the single void case. In the main figure the void fraction $f$ has
been scaled with the exponential. The circles point where the
dynamical ILD's cross the line ILD$=0.5$ in Fig.~\ref{fig2}.}
\label{fig4}
\end{figure}

We have identified the onset of coalescence, but it is also
interesting to examine the void growth following the onset of void
interactions but prior to the actual coalescence.  How does this
differ from the exponential growth of an isolated void
\cite{belak,prb}?  In order to analyze the correlated growth, we have
factored out the non-interacting growth rate, $\exp(200\epsilon)$,
from $f$ in the plot in Fig.~\ref{fig4}.  The factor of 200 in the
exponential is derived from the single void case, as indicated in the
inset of the figure.  The void growth data for ILD$_0$=4.6 and 1.8
coincide with the single void curve.  The void growth rate with
smaller ILD$_0$'s reach their asymptotic growth rate earlier.  In the
figure we have drawn as circles the void size values, where the
dynamic ILD's cross the line ILD=0.5 in Fig.~\ref{fig2}. As can be
seen from figure, there is no marked change in the void volume
behavior when the voids start to interact.

We have also performed a series of simulations of a fixed void in
varying box size in order to find the coalescence process of the void
with its (six) periodic image(s), similar to the manner in which some
continuum calculations of coalescence have been done.  The details
will be reported elsewhere~\cite{unpub}, but it is worth noting, that
the behavior is opposite to the results above in an important way.
The smaller the box size, and hence the smaller the ILD, the later the
void starts to grow.

To summarize, interaction and coalescence of two voids in copper under
tension have been simulated in multi-million-atom MD simulations. The
effect of interactions between voids has been quantified by the
increased reduction-rate of their separation and the movement of their
centers.  The interaction between the voids can be also measured in
their shape changes~\cite{prb} as will be done
elsewhere~\cite{unpub}. The critical inter-void ligament distance has
been found to be close to one void radius, independent of the
strain-rate or the starting separation ILD$_0$.  The onset of
coalescence occurs at the point that the plastic zones surrounding the
voids first interact strongly.

\begin{acknowledgments}

This work was performed under the auspices of the US Dept.\ of Energy by 
the Univ.\ of Cal., Lawrence Livermore National Laboratory, under
contract no.\ W-7405-Eng-48.

\end{acknowledgments}





\begin{thebibliography}{200}

\bibitem{Curran} 
D.\ R.\ Curran, L.\ Seaman, and D.\ A.\ Shockey,
Phys.\ Rep.\ {\bf 147}, 253 (1987).

\bibitem{McClintock}
F.\ A.\ McClintock
in {\it Metallurgical Effects at High Strain Rates},
edited by R.\ W.\ Rhode, B.\ M.\ Butcher, and J.\ R.\ Holland
(Plenum Press, New York, 1973).

\bibitem{McClintock2}
F.\ A.\ McClintock,
J.\ Appl.\ Mech.\ {\bf 6}, 363 (1968).

\bibitem{RiceTracey}
J.\ R.\ Rice and D.\ M.\ Tracey,
J.\ Mech.\ Phys.\ Solids {\bf 17}, 201 (1969).

\bibitem{Gurson} 
A.\ L.\ Gurson,
J.\ Eng.\ Mater.\ and Tech.\ {\bf 99}, 2 (1977).

\bibitem{tvergaard} 
J.\ Koplik and A.\ Needleman,
Int.\ J.\ Solids Structures {\bf 24}, 835 (1988).

\bibitem{belak} 
J.\ Belak, 
in {\it Shock Compression of Condensed Matter},
edited by Schmidt {\it et al.}
(American Institute of Physics, New York, 1997).

\bibitem{rudd} 
R.\ E.\ Rudd and J.\ Belak,
Comp.\ Mat.\ Science, 
{\bf 24}, 148 (2002).

\bibitem{prb}
E.\ T.\ Sepp\"al\"a, J.\ Belak, and R.\ E.\ Rudd, 
Phys.\ Rev.\ B {\bf 69}, 134101 (2004).

\bibitem{koss}
J.\ P.\ Bandstra and  D.\ A.\ Koss,
Mat.\ Sci.\ Eng.\ A {\bf 319-321}, 490 (2001);
A.\ B.\ Geltmacher, D.\ A.\ Koss, P.\ Matic, and M.\ G.\ Stout,
Acta Mater.\ {\bf 44}, 2201 (1996);
D.\ M.\ Goto and D.\ A.\ Koss,
Scripta Mater.\ {\bf 35}, 459 (1996);
P.\ E.\ Magnusen, D.\ J.\ Srolovitz, and D.\ A.\ Koss,
Acta Metall.\ Mater.\ {\bf 38}, 1013 (1990);
P.\ E.\ Magnusen, E.\ M.\ Dubensky, and D.\ A.\ Koss,
Acta Metall.\ {\bf 36}, 1503 (1988);
V.\ Jablokov, D.\ M.\ Goto, and D.\ A.\ Koss,
Metall.\ and Mater.\ Trans.\ A {\bf 32}, 2985 (2001).

\bibitem{somerday}
B.\ P.\ Somerday, P.\ D.\ Pattillo II, M.\ F.\ Horstemeyer, and M.\ I.\ Baskes,
Mat.\ Res.\ Soc.\ Symp.\ Proc.\ Vol.\ {\bf 578}, 333 (2000).

\bibitem{Eshelby}
J.\ D.\ Eshelby, 
Proc. Royal. Soc. A {\bf 252}, 561 (1959).

\bibitem{Brown} L.\ M.\ Brown and J.\ D.\ Embury 
in {\it Proceedings of the third International Conference on 
Strength of Metals and Alloys} (Institute of Metals, London, 1973).

\bibitem{horstemeyer}
M.\ F.\ Horstemeyer, M.\ M.\ Matalanis, A.\ M.\ Sieber, M.\ L.\ Botos,
Int.\ J.\ Plasticity {\bf 16}, 979 (2000).

\bibitem{allen}
M.\ P.\ Allen and D.\ J.\ Tildesley,
{\it Computer Simulations of Liquids}
(Oxford University Press, Oxford, 1987).


\bibitem{oh1} 
D.\ J.\ Oh and R.\ A.\ Johnson, 
J.\ Mater.\ Res.\ {\bf 3}, 471 (1988).

\bibitem{hoover}
W.\ G.\ Hoover,
Phys.\ Rev.\ A {\bf 31}, 1695 (1985).

\bibitem{parrinello}
M.\ Parrinello and A.\ Rahman,
J.\ Appl.\ Phys.\ {\bf 52}, 7182 (1981).


\bibitem{hamilton} 
C.\ L.\ Kelchner, S.\ J.\ Plimpton, and J.\ C.\ Hamilton,
Phys.\ Rev.\ B {\bf 58}, 11085 (1998).

\bibitem{unpub}
E.\ T.\ Sepp\"al\"a, J.\ Belak, and R.\ E.\ Rudd, 
unpublished.


\end{thebibliography}
\end{document}